\documentclass{ws-ijmpe}

\begin{document}

\markboth{Christof Roland, G\'abor I. Veres, Kriszti\'an Krajcz\'ar}
{Simulation of jet quenching observables in Heavy Ion collisions at the LHC}

\catchline{}{}{}{}{}
%%%%%%%%%%%%%%%%%%%%%%%%%%%%%%%%%%%%%%%%%%%%%%%%%%%%%%%%%%%%%%%%%%%%

\title{SIMULATION OF JET QUENCHING OBSERVABLES\\
IN HEAVY ION COLLISIONS AT THE LHC}

\author{\footnotesize CHRISTOF ROLAND\footnote{address: CERN, Switzerland}}

\address{Laboratory for Nuclear Science, Massachussetts Institute of Technology\\
77 Massachusetts Avenue, Cambridge, MA 02139-4307, USA
%\footnote{State completely without abbreviations, the
%affiliation and mailing address, including country.}
\\
christof.roland@cern.ch}

\author{G\'ABOR I. VERES, KRISZTI\'AN KRAJCZ\'AR}

\address{Department of Atomic Physics, E\"otv\"os Lor\'and University\\
P\'azm\'any P\'eter s\'et\'any 1/A, Budapest, H-1117, Hungary\\
vg@ludens.elte.hu, nirnaeth@monornet.hu}

\author{for the CMS Heavy Ion Group}

\maketitle

\begin{history}
\received{(received date)}
\revised{(revised date)}
%\accepted{(Day Month Year)}
%\comby{(xxxxxxxxxx)}
\end{history}

\begin{abstract}
Large transverse momentum jets provide unique tools to study
dense QCD matter in high-energy heavy-ion collisions.
Results from RHIC on suppression of high transverse momentum
particles in Au+Au collisions indicate a significant energy
loss of leading partons in the dense and strongly interacting
matter formed in these collisions. 
The LHC will collide Pb ions at $\sqrt{s_{_{NN}}}$=5500~GeV, where the 
cross section of hard scattering will increase 
dramatically. Large production rates, the large acceptance of
the CMS calorimeters and tracking system, combined with the 
capability of triggering on jets, will extend
the transverse momentum reach of charged 
particle spectra and nuclear modification factors up to 
$p_T>200$ GeV/$c$.
\end{abstract}

%----------------------------------------------------------------------
\section{Introduction}

The abundance of high $Q^2$ processes at LHC energies will provide large 
samples of high $E_{\rm T}$ jets, large $p_{\rm T}$ hadrons, and jets produced 
opposite to gauge bosons ($\gamma^{\star}, Z$).\cite{brandt}
The strong interest in 
these observables in heavy-ion collisions stems from the concept that high 
$E_{\rm T}$ quark and gluon jets can be used to probe the hot and dense medium 
produced in the collision, because they are affected by the properties of 
the medium as they propagate through this dense environment.
Partons with high transverse momentum are predicted to suffer radiative 
and collisional energy loss in the created medium, suppressing the yield 
of jets and particles found with high transverse energy in a heavy-ion 
collision, compared to the p+p collision case (see e.g. \cite{salgado}).

Early results obtained at RHIC indeed showed
suppression of the hadron yield at $p_{\rm T}>3$~GeV/$c$ and the
disappearance of back-to-back correlations of high-$p_{\rm T}$ particles.\cite{jetq_rhic}
These indirect measurements of jet properties suggest a significant
in-medium energy loss of fast partons, which
will be experimentally accessible at the LHC also by observing fully formed and 
reconstructed jets.

The performance of the CMS detector for Pb+Pb events was extensively studied in
full simulations with realistic assumptions for
particle multiplicity, jet and hadron spectra.\cite{lokhtin}
The charged particle reconstruction capabilities using the
CMS Silicon Tracking System are evaluated using a full detector simulation,
assuming a charged particle density in central Pb+Pb collisions
of ${\rm d}N_{\rm ch}/{\rm d}y=3200$. In this
high multiplicity environment, an algorithmic tracking efficiency of about 80\%
is achieved, with less than 5\% fake track rate
for $p_{\rm T}>1$~GeV/$c$ and excellent momentum
resolution, $\Delta p_{\rm T}/p_{\rm T}<1.5\%$ (for $p_{\rm T}<100$~GeV/$c$).~\cite{cmshitrk}

The High Level Trigger of the CMS data acquisition system is sufficiently 
powerful to allow the inspection of
all minimum bias Pb+Pb events individually, where 
the full event information will be available for the trigger decision.\cite{hltnote}
Jets are reconstructed in the calorimeters using an iterative
cone algorithm, which is modified to subtract the underlying soft
background, and can be included in the HLT. 
The lower limit of transverse energy needed
for efficient and clean reconstruction is about 50 GeV.
The energy resolution for jets with 100 GeV transverse energy at
$\eta\approx 0$ is about 16\%.\cite{kodolova} 
%An important application to combine
%the jet trigger and particle reconstruction capabilities in CMS will be discussed
%in the following sections.

%--------------------------------------------------------------------
\section{Nuclear Modification Factor}

The nuclear modification factor, $R_{AA}$, and the central to peripheral
ratio, $R_{CP}$, provide quantitative information
on the amount and energy-dependence of the energy-loss of
hard-scattered partons that traverse the
medium created in the heavy-ion collision.
They are defined as:

\begin{equation}
R_{AA}=\frac{\sigma_{pp}^{\rm inel}}{\langle N_{\rm coll}\rangle}
\frac{{\rm d}^2N_{AA}/{\rm d}p_{\rm T}{\rm d}\eta}{{\rm
    d}^2\sigma_{pp}/{\rm d}p_{\rm T}{\rm d}\eta}\quad {\rm and} \quad
R_{CP}=\frac{\langle N_{\rm coll}^{\rm periph}\rangle}{\langle N_{\rm coll}^{\rm central}\rangle}
\frac{{\rm d}^2N_{AA}^{\rm central}/{\rm d}p_{\rm T}{\rm d}\eta}{{\rm
      d}^2N_{AA}^{\rm periph}/{\rm d}p_{\rm T}d\eta}, \label{one}
\end{equation}

\noindent
where $\langle N_{\rm coll}\rangle$ is the number of binary
nucleon-nucleon collisions in the geometrical Glauber picture, averaged over
the events that belong to a given centrality class.

$R_{AA}$ quantifies the suppression (or enhancement) of hadron production
with respect to p+p collisions.
The invariant cross section of charged particles at high $p_{\rm T}$ is
expected to scale with $\langle N_{\rm coll}\rangle$ if no nuclear
effects take place, in which case the value of $R_{AA}$ at high $p_{\rm T}$ would
be unity.
$R_{CP}$ does not require a p+p reference,
as it compares central and peripheral heavy-ion collisions. %However,
It is not equivalent to $R_{AA}$, since even the most peripheral heavy-ion
collisions are influenced by nuclear effects.

The $R_{AA}(p_{T})$ and $R_{CP}(p_{T})$ for charged hadrons thus provide
important information on the properties of the
created medium, like the initial gluon rapidity density, ${\rm d}N_g/{\rm d}y$, or the transport
coefficient, $\langle\hat{q}\rangle$.\cite{salgado}

\section{Triggering on Jets}

\begin{figure}[t]
\centerline{\psfig{file=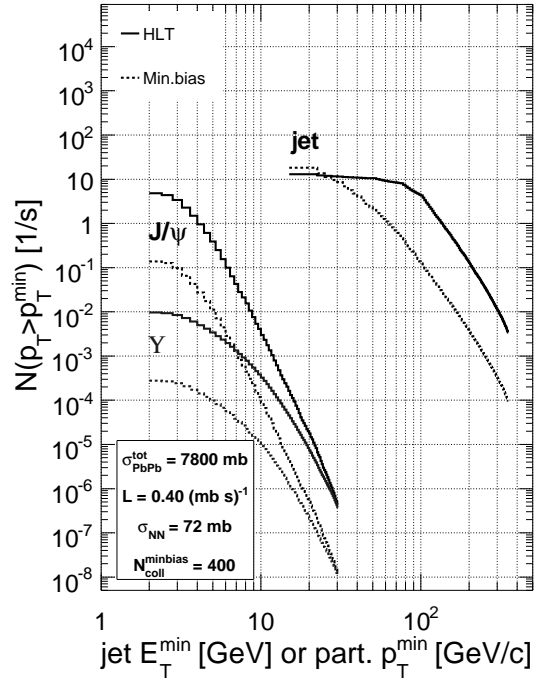,width=70mm}}
\vspace*{8pt}
\caption{\label{rates}
Event rate of various hard processes stored on tape above
a certain $E_{\rm T}$ ($p_{\rm T}$) at the design Pb+Pb luminosity.
The minimum bias and the HLT data taking modes are compared.}
\end{figure}

For both $R_{AA}$ and $R_{CP}$,
triggering on jets
will be essential to extend the measurable $p_{\rm T}$ range.
Using the CMS calorimeter towers, it is possible to quickly
reconstruct these energetic jets with a good energy resolution.
The jet reconstruction algorithm can be included in the High Level 
Trigger, and will record events containing a high 
energy jet with high efficiency.\cite{hltnote} 
The increase in the number of recorded jets using 
the jet trigger is illustrated in Fig.~\ref{rates}.

Charged hadrons at $p_{\rm T}>20-30$~GeV/$c$ originate mainly from the fragmentation of high
$E_{\rm T}$ jets: they are typically the leading hadrons of the energetic jets.
The jet trigger is useful to collect sufficiently large jet
statistics to study fragmentation functions, jet correlations,
and other observables, among which we will present here the charged particle nuclear
modification factors.
To provide adequate and fast Monte Carlo tools to simulate jet quenching,
the Monte Carlo event generator HYDJET (HYDrodynamics plus JETs) has been
developed and is used to produce heavy-ion collisions
at LHC energies.\cite{lokhtin,hydjet}
Final state particles in nuclear collisions from HYDJET are
obtained as a superposition of soft hydro-type particle production and
multiple hard parton-parton collisions.
%This model is capable of reproducing hadron momentum spectra
%and main features of the jet quenching pattern in heavy ion collisions
%observed at RHIC energies: the $p_{\rm T}$ dependence of the nuclear 
%modification
%factor and the suppression of azimuthal back-to-back
%correlations.\cite{lokhtin}

\section{Results}

Within the 15\% of the full bandwidth assigned to minimum bias events,
13.5 million events are expected to be taken in one month. 
The jet triggers with 50, 75 and 100 GeV $E_T$ thresholds will be able to 
sample 0.35, 1.9 and 3.9 billion events, respectively. It is possible to 
generate the amount of minimum bias events 
with our generator-level tools. However, triggering enhances the 
number of jets at high $E_{\rm T}$ by more than two orders of magnitude. 
Thus, instead of generating one hundred
times more minimum bias heavy-ion events to conduct our study, we have
implemented a ``trigger'' at the generator level.
This way, we only store simulated events which
are likely to produce a large $E_{\rm T}$ jet when the jet
finder is run on the calorimeter towers.
Thus, more events are sampled than stored and 
the necessary scaling factors
between the different trigger samples are not immediately available.

\begin{figure}[t]
\centerline{\psfig{file=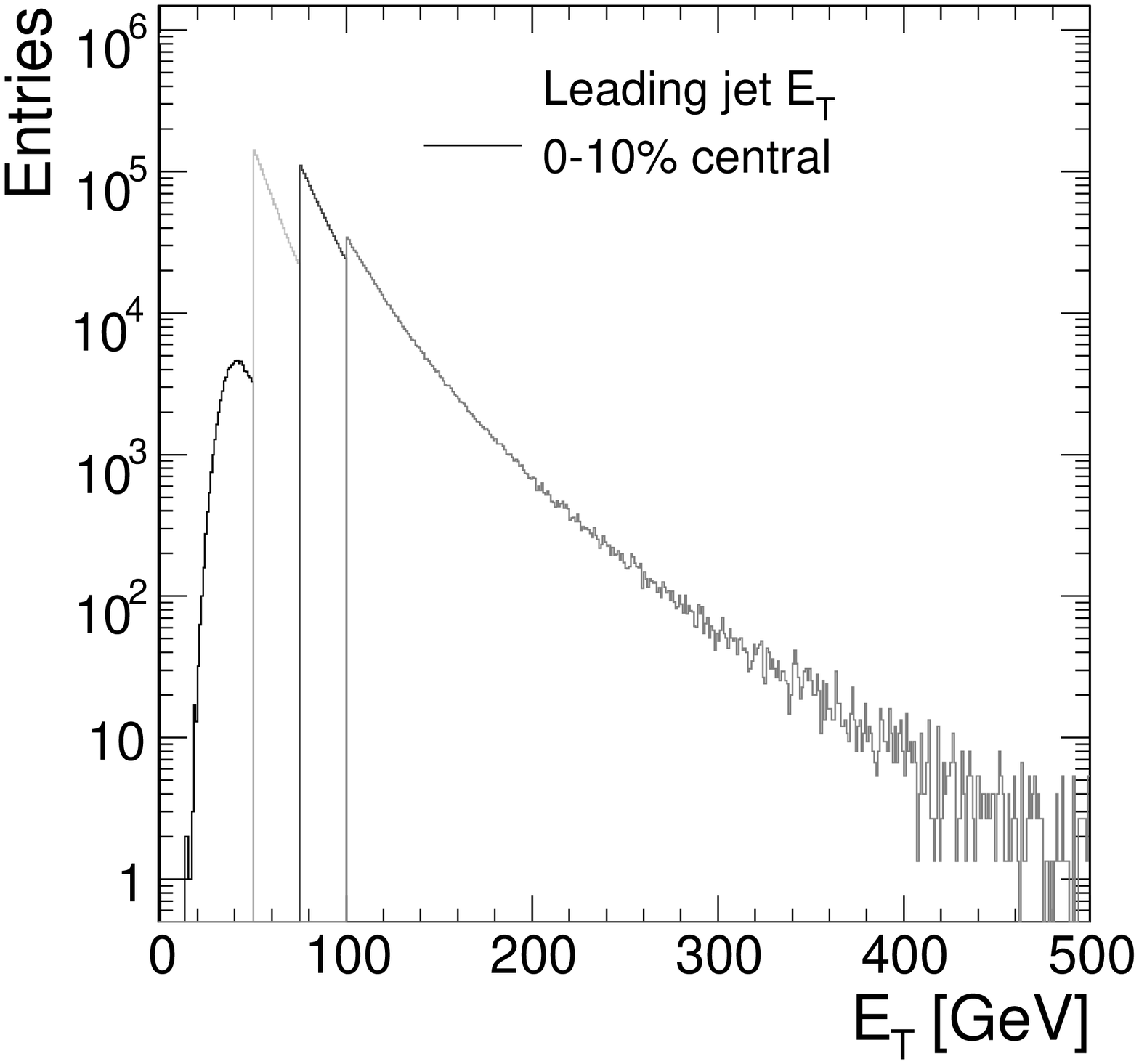,width=6cm}
\psfig{file=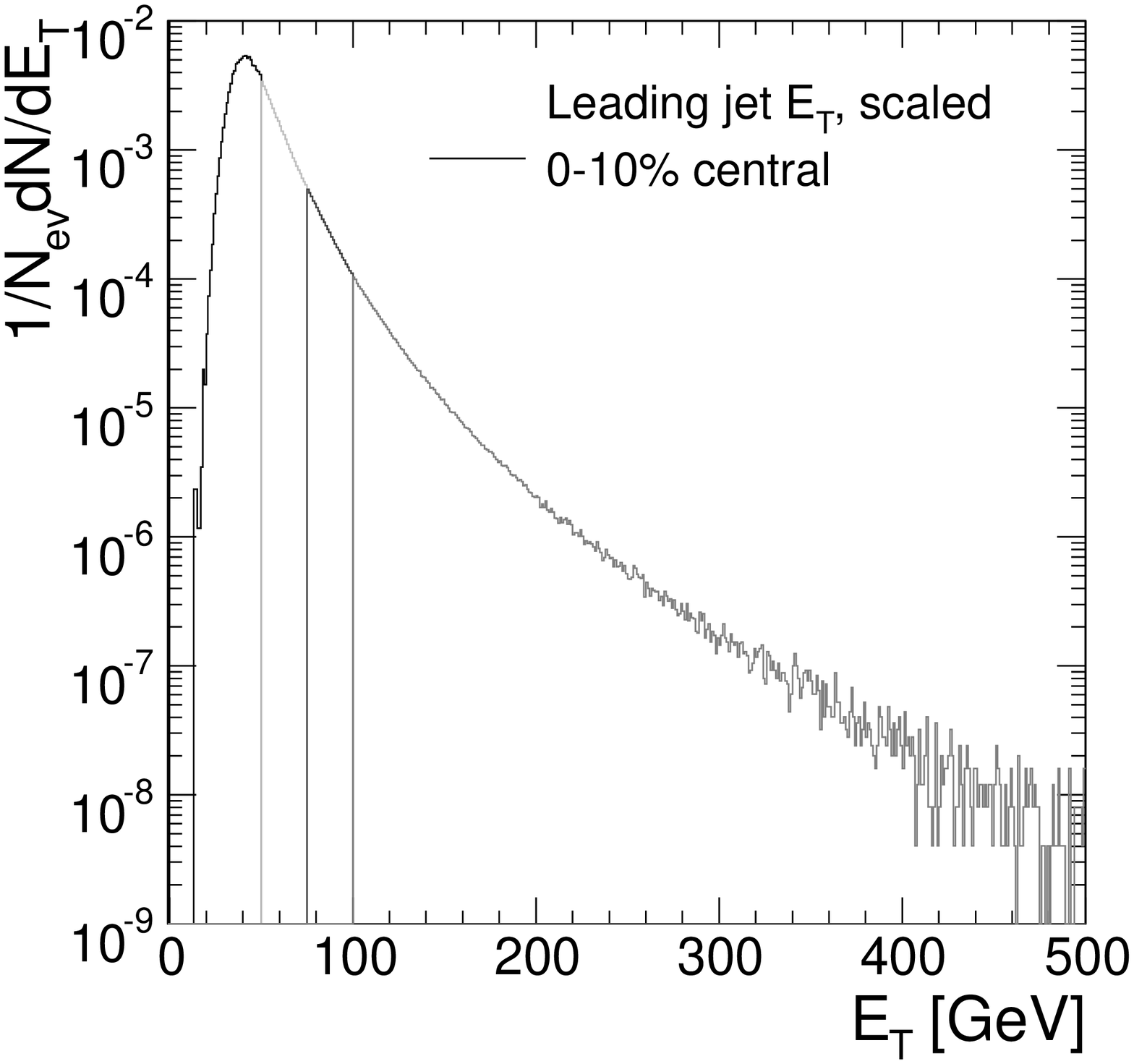,width=6cm}}
\vspace*{-1mm}
\caption{\label{jetettrig}
Left panel: leading jet $E_{\rm T}$ distributions for minimum bias (black),
and triggered simulated data samples, with $E_{\rm T}$ thresholds of 50, 75 and 100 GeV. Right panel: the 
same distributions sliced into the 0--50, 50--75, 75--100 and 100--300 GeV
intervals, and scaled by the appropriate factors to get back the
non-triggered distribution, with significantly higher statistics.}
\end{figure}

\begin{figure}[t]
\centerline{\psfig{file=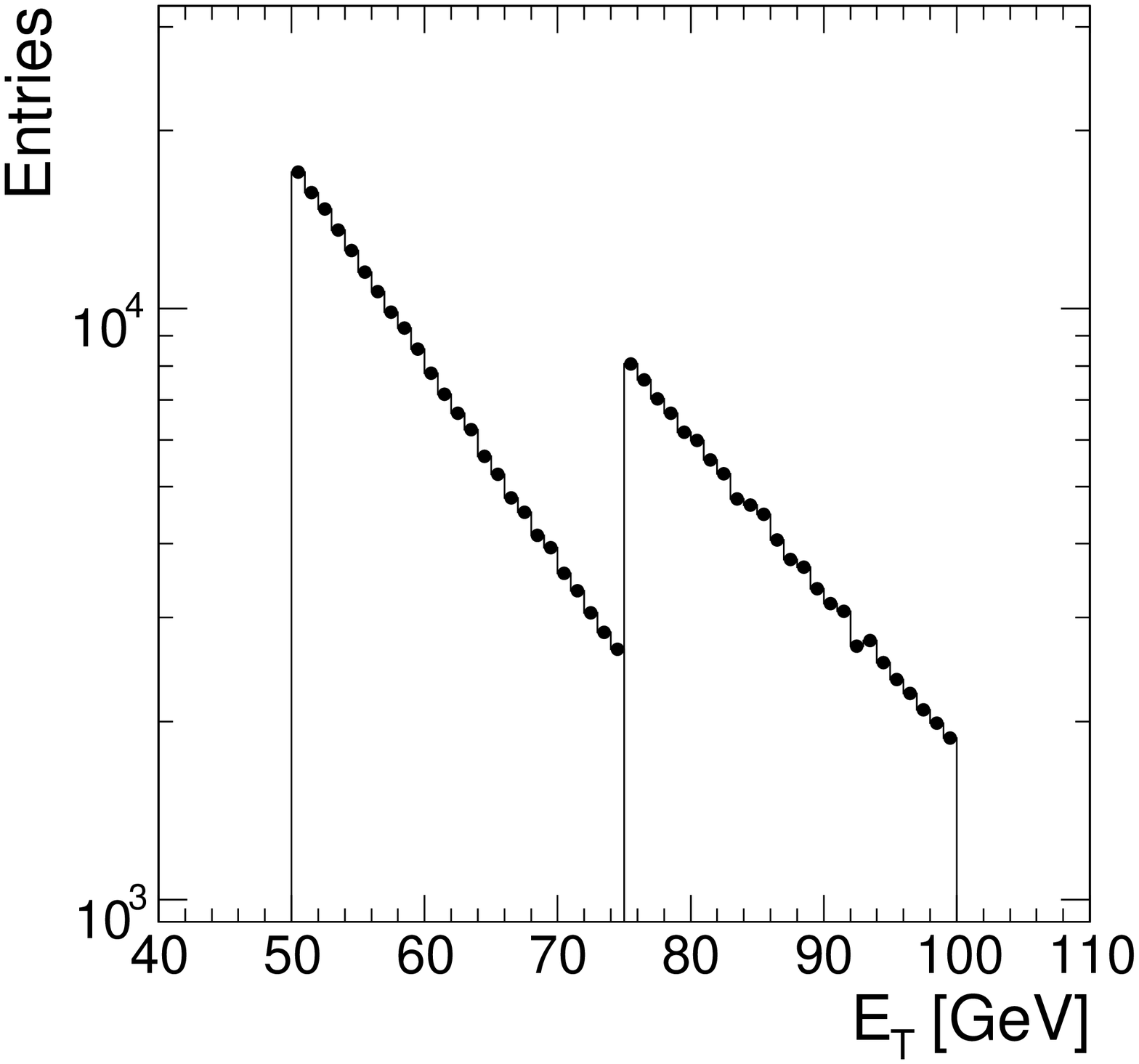,width=6cm}
\psfig{file=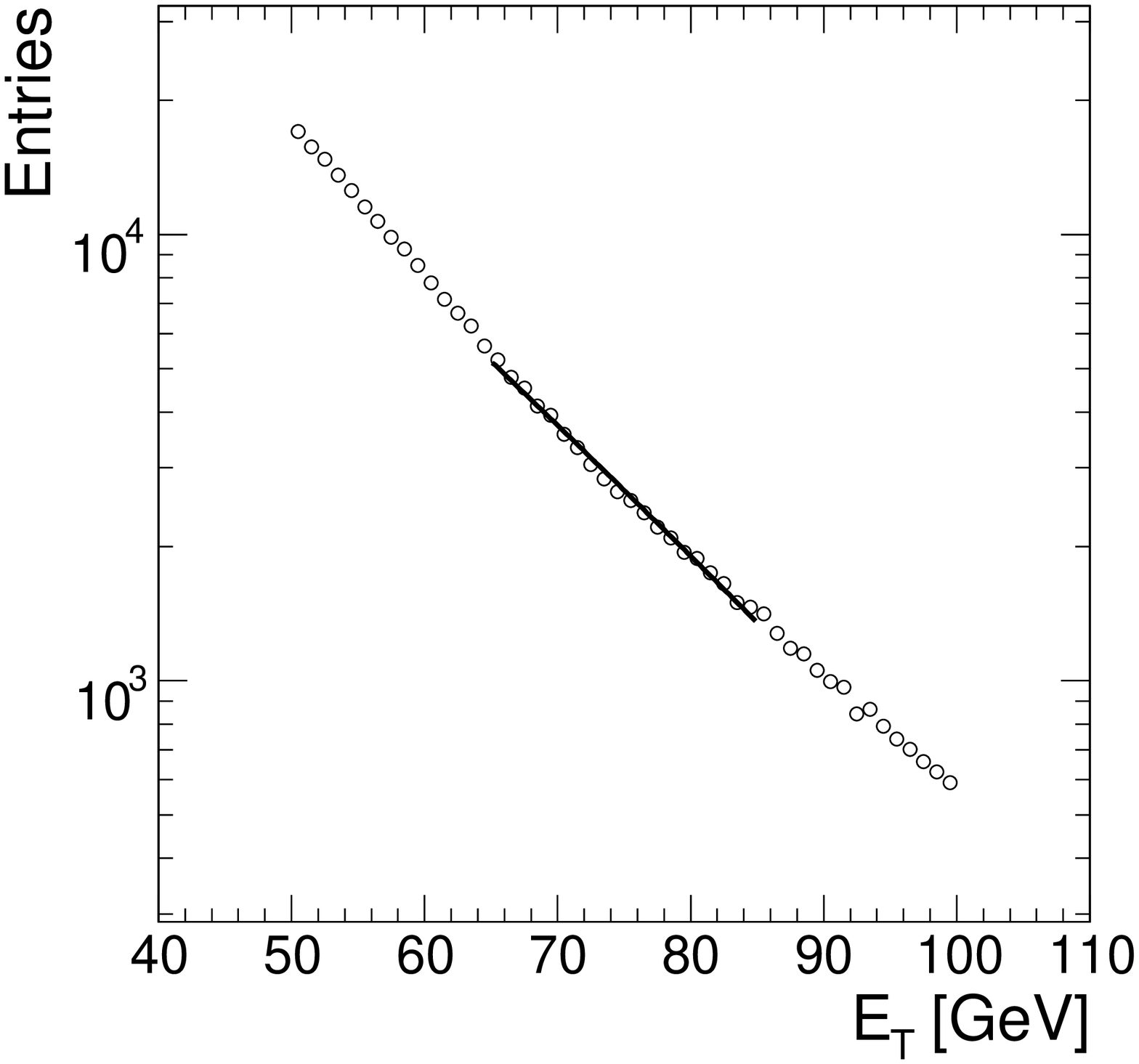,width=6cm}}
\vspace*{-0.5mm}
\caption{\label{fitproc}
Left panel: leading jet $E_{\rm T}$ distributions for two consecutive
triggered data samples. Right panel: the same distributions scaled
with the factor obtained in the merging procedure.}
\end{figure}

Figure~\ref{jetettrig} shows %illustrates the following procedure.
the distribution of the highest $E_{\rm T}$ jet within $|\eta|<2$ per event,
the ``leading'' jet, % is plotted 
for minimum bias
(black histogram), and for jet-triggered events with 50, 75 and
100 GeV thresholds (grey histograms).
The scaling factors between consecutive data sets are determined by joining
them with scaling factors determined by fitting the combined leading jet
$E_{T}$ spectrum with a power law in the joining regions.  This way,
the optimal scaling factors
can be determined from the data distribution, without any prior
assumptions on the
spectrum. This is illustrated in Fig.~\ref{fitproc}: the left panel
shows the data sets before merging while the right panel
shows the same data sets after scaling, together with the
power law fit.
The minimum bias $E_{\rm T}$ spectrum is recovered with 
significantly increased statistics at high $E_{\rm T}$.

Using this method to merge data sets with different
thresholds, triggered data sets were generated
with the number of jets with $E_{\rm T}$
above the thresholds expected from one month of Pb+Pb
data taking at design luminosity. The statistical errors 
on the charged particle $p_{\rm T}$ spectrum in the merged data set
reflects, hence, the real experimental situation
after one month of data taking, with the four different trigger thresholds.

\begin{figure}[t]
\centerline{\psfig{file=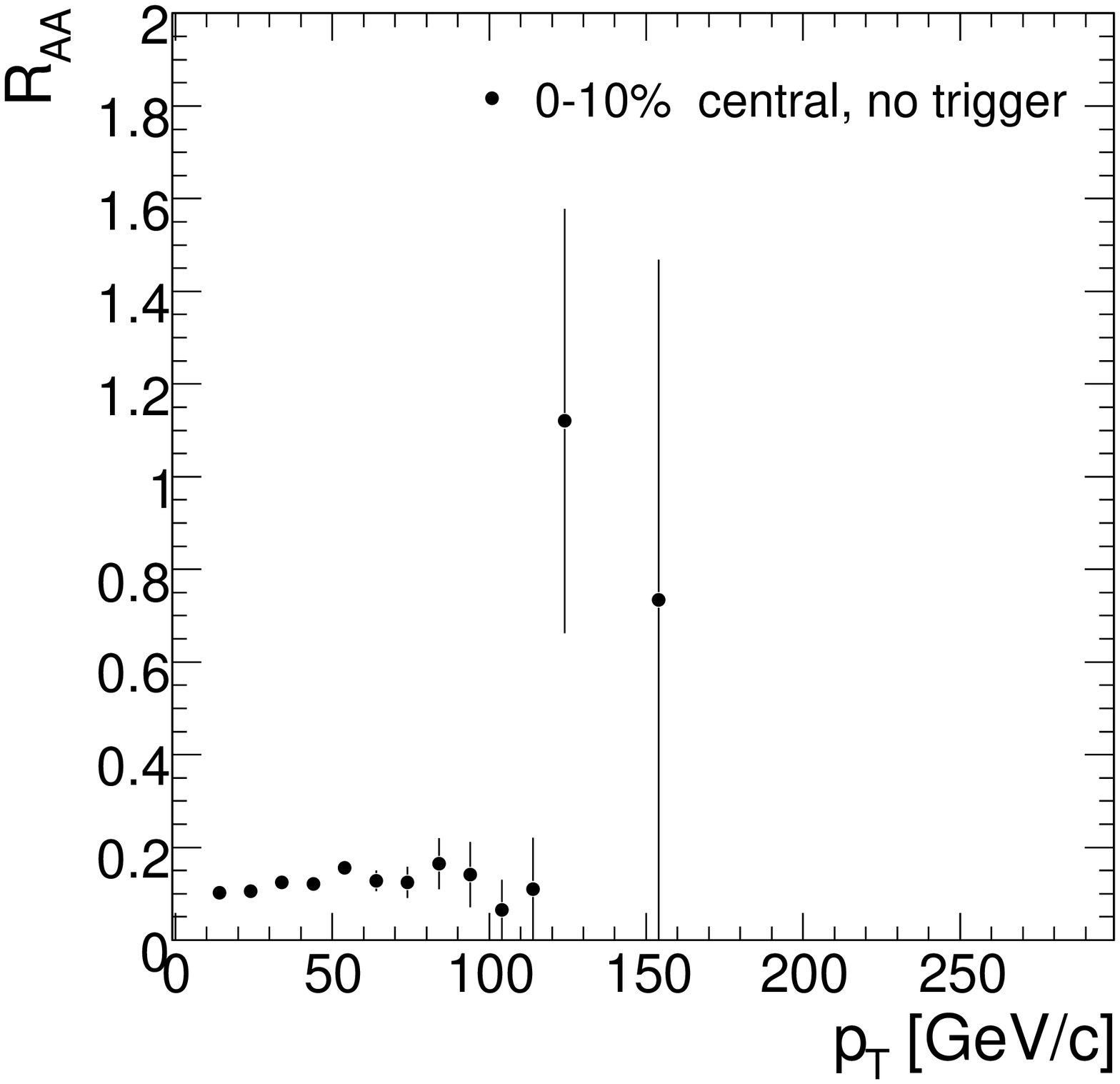,width=6cm}
\psfig{file=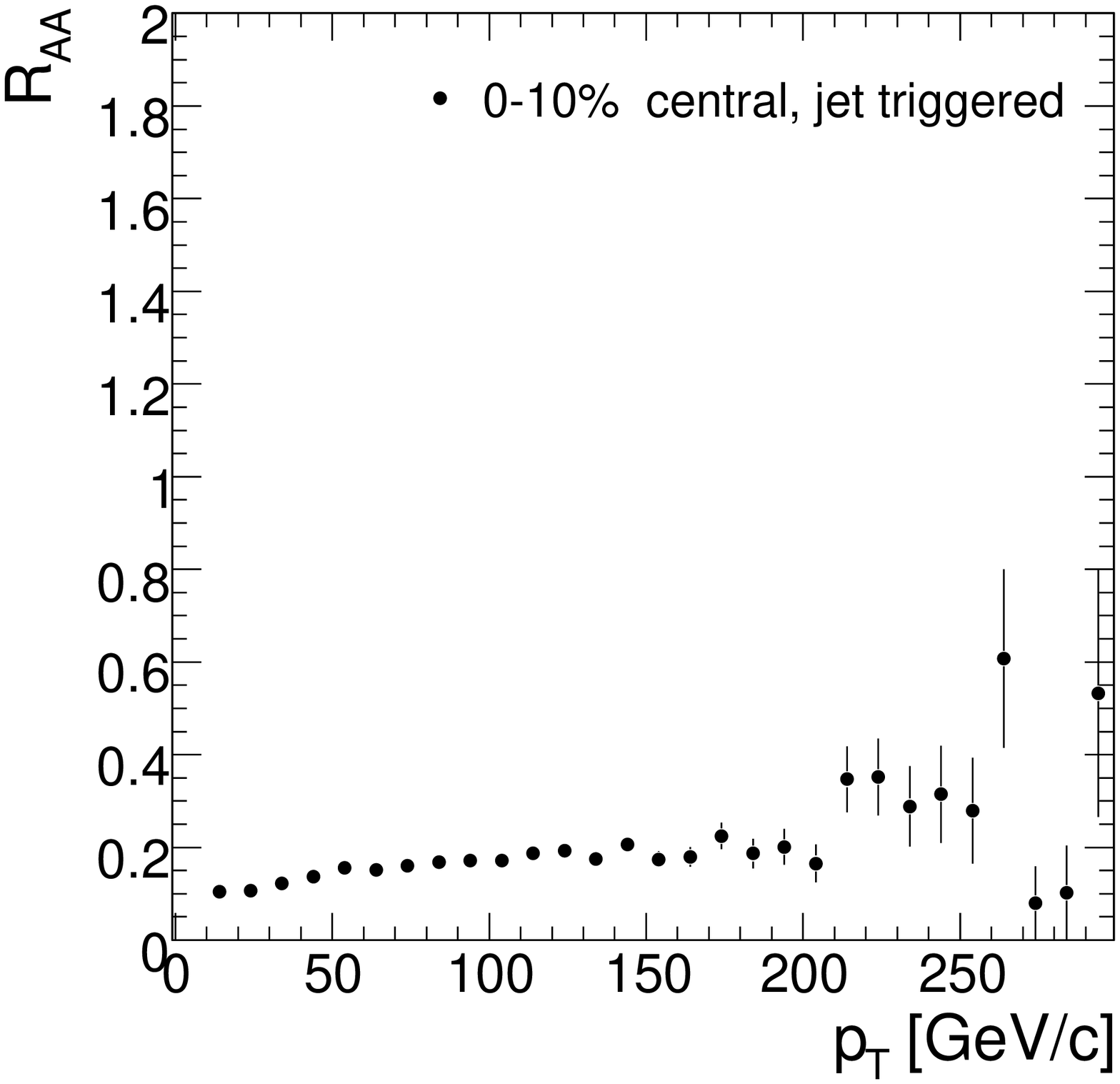,width=6cm}}
\vspace*{-0.5mm}
\caption{\label{raa}
The nuclear modification factor $R_{AA}$ as a function of $p_{\rm T}$ for charged
particles, for minimum bias data (left) and for data triggered
on high-$E_{\rm T}$ jets (right), for one month of data taking.}
\end{figure}

\begin{figure}[t]
\centerline{\psfig{file=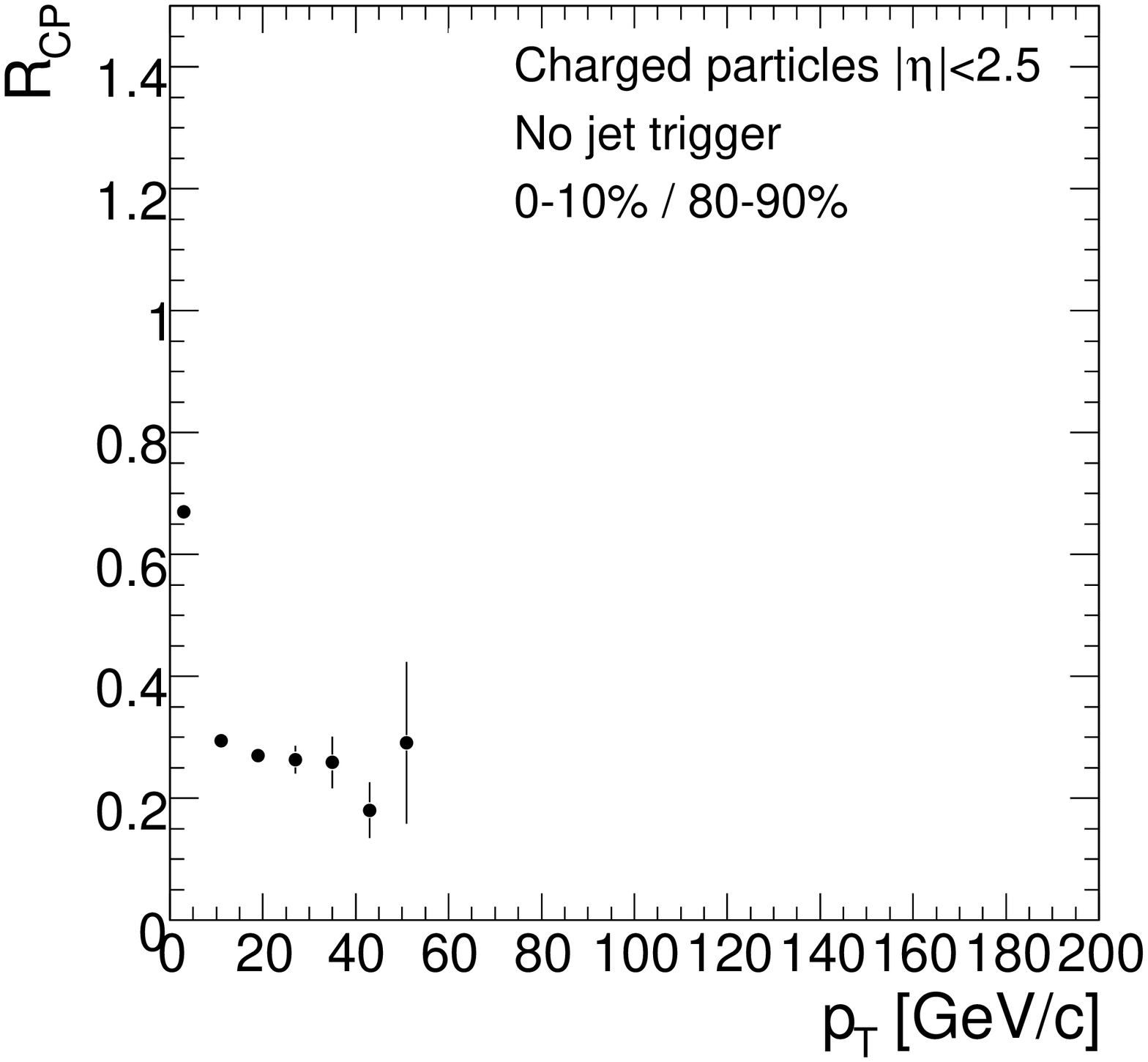,width=6cm}
\psfig{file=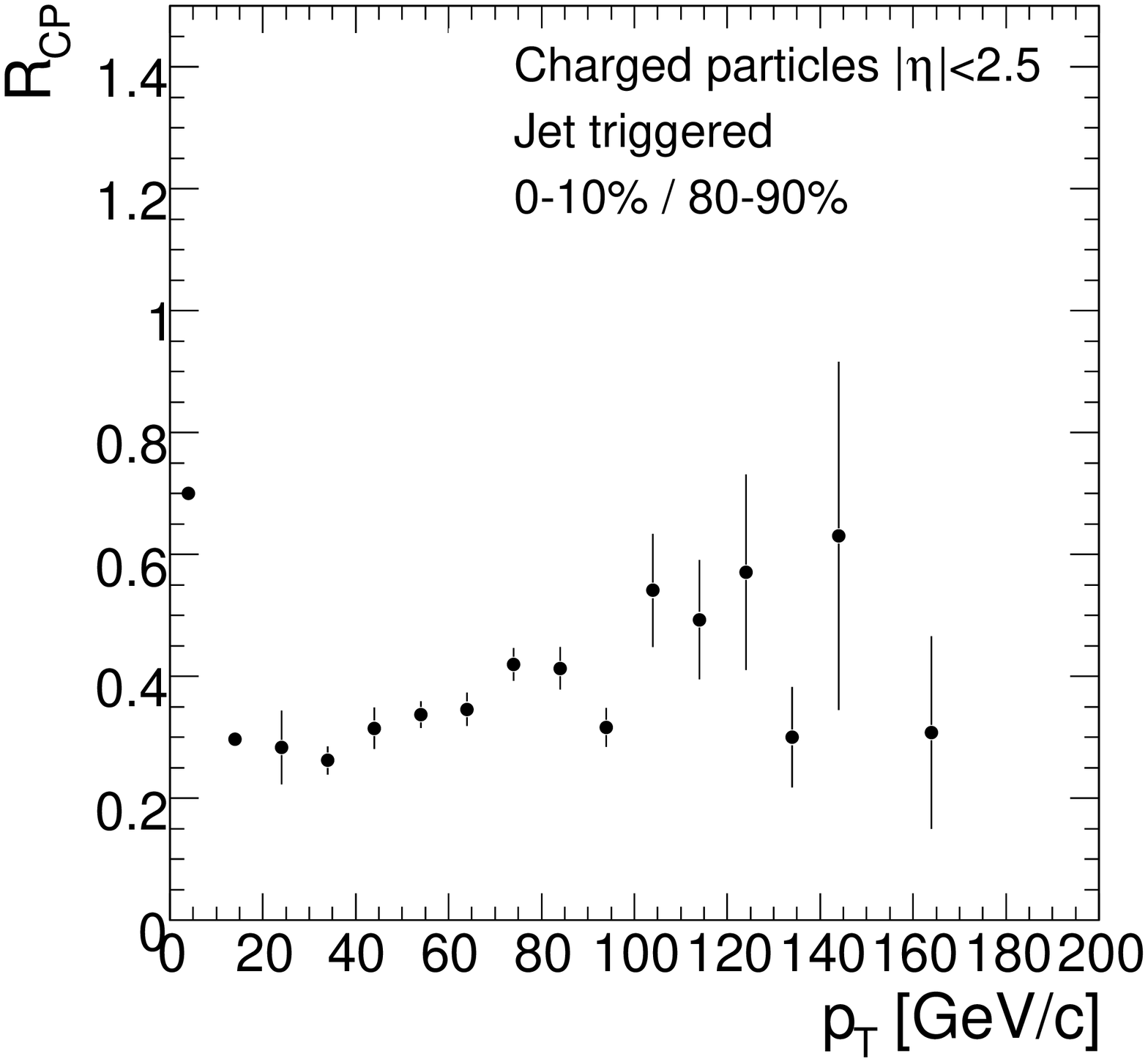,width=6cm}}
\vspace*{8pt}
\caption{\label{rcp}
The central-to-peripheral ratio, $R_{CP}$, as a function of $p_{\rm T}$, for charged
particles, for minimum bias data (left) and for data triggered
on high-$E_{\rm T}$ jets (right), for one month of data taking.}
\end{figure}

The obtained $R_{AA}$ is shown in Fig.~\ref{raa}.  
In the present study, the PYTHIA event generator was
used to simulate the p+p reference for $R_{AA}$.
%Another possibility is to measure 
The $R_{CP}$ ratio, which uses
peripheral Pb+Pb collisions as reference, instead of p+p collisions,
is shown in Fig.~\ref{rcp}, % shows the result for the $p_{\rm T}$ reach of $R_{CP}$
for one month of data taking at nominal luminosity.
Comparing the results for the minimum bias data (left panels of
Figs.~\ref{raa} and~\ref{rcp}) to the results for the jet
triggered data (right panels), we see that triggering on jets
significantly extends the
$p_{\rm T}$ range of $R_{AA}$ and $R_{CP}$,
from $\sim$\,100 to $\sim$\,250~GeV/$c$ in $R_{AA}(p_{\rm T})$ and 
from $\sim$\,50 to $\sim$\,150~GeV/$c$ in $R_{CP}(p_{\rm T})$.

\section*{Acknowledgements}

The authors are indebted to M. Ballintijn, C. Loizides, I.~P.~Lokhtin, S.~V.~Petrushanko, 
G. Roland, L. I. Sarycheva, A. M. Snigirev, I. N. Vardanyan, B. Wyslouch, and to 
the CMS Heavy Ion group for valuable discussions, and to all contributors of the HIROOT 
analysis tool. Two of us (G.I.V. and K.K.) thanks the support of the 
Hungarian Scientific
Research Fund (T 048898 and F 049823).


\begin{thebibliography}{8}
\bibitem{brandt} D. Brandt, {\it LHC Project Report} (2000) 450.   %1
\bibitem{salgado} C. A. Salgado and U. A. Wiedemann, {\it Phys. Rev.} {\bf D68} (2003) 014008.   %2
\bibitem{jetq_rhic} 
 K. Adcox {\it et al.} (PHENIX), {\it Phys. Lett.} {\bf B561} (2003) 82-92,\\
 B. B. Back {\it et al.} (PHOBOS), {\it Phys. Rev. Lett.} {\bf 91} (2003) 072302,\\
 J. Adams {\it et al.} (STAR), {\it Phys. Rev. Lett.} {\bf 91} (2003) 072304,\\
 B. B. Back {\it et al.} (PHOBOS), {\it Phys. Lett.} {\bf B578} (2004) 297-303,\\
 B. B. Back {\it et al.} (PHOBOS), {\it Phys. Rev. Lett.} {\bf 94} (2005) 082304
\bibitem{lokhtin} I. P. Lokhtin and A. M. Snigirev, {\it Eur. Phys. J.} {\bf C45} (2006) 211.   %9
\bibitem{cmshitrk} C. Roland, {\it CMS Note} 2006/031.   %10
\bibitem{hltnote} M. Ballintijn, C. Loizides and G. Roland,
{\it ArXiv:nucl-ex/0702041}, "CMS Physics TDR Addendum: High Density QDC
with Heavy-Ions", CERN-LHCC-2007, to be submitted
\bibitem{kodolova} O. Kodolova {\it et al.}, {\it CMS Note} 2006/050.   %13
\bibitem{hydjet} {\it http://cern.ch/lokhtin/hydro/hydjet.html}   %15
\end{thebibliography}
\end{document}